# What's in People's Digital File Collections?

**Jesse David Dinneen**
Victoria University of
Wellington, New Zealand
jesse.dinneen@vuw.ac.nz

**Charles-Antoine Julien**
McGill University
Montreal, Canada
charles.julien@mcgill.ca

**ABSTRACT**
Thoughtfully designing services and rigorously testing software to support personal information management (PIM) requires understanding the relevant collections, but relatively little is known about what people keep in their file collections, especially personal collections. Complementing recent work on the structure of 348 file collections, we examine those collections' contents, how much content is duplicated, and how collections used for personal matters differ from those used for study and work. Though all collections contain many images, some intuitively common file types are surprisingly scarce. Personal collections contain more audio than others, knowledge workers' collections contain more text documents but far fewer folders, and IT collections exhibit unusual traits. Collection duplication is correlated to collections' structural traits, but surprisingly, not to collection age. We discuss our findings in light of prior works and provide implications for various kinds of information research.

**ASIS&T THESAURUS**
1577 Personal files; 2988 Personal documents; 1469 Personal collections; 1524 Interfaces

**INTRODUCTION**
The task of managing digital files, also called file management (FM), is ubiquitous in computing: at home and at work, people create, download, copy, name, rename, organise (or leave unorganised), delete (or not), share, navigate to, and search for digital files and folders (Dinneen & Julien, 2019). The result of such activities are collections of files and folders people personally manage for personal, professional, and educational purposes. Such collections are very large, typically containing tens and even hundreds of thousands of files (Dinneen, Julien, & Frissen, 2019). But what is in these collections? What are people storing?

To our knowledge, only five recent (e.g., since 1999) studies have examined the contents of people's file collections, though none have examined *strictly* personal collections (i.e., those not used in work or study), and while several studies of personal information management (PIM) have suggested occupational factors likely determine users' behaviour and collections, empirical data for comparing such collections has been scarce. Further, though two studies (Henderson, 2011; Hicks, Dong, Palmer, & McAlpine, 2008) found a surprising amount of duplication of contents in work collections, the causes of duplication remain unknown.

As knowledge about the contents of file collections is limited, so too is our ability to model and understand them, test and compare relevant software improvements, and ultimately support their management. For example, it is unclear what kinds of files should comprise representative test collections to be used in evaluating PIM prototype software, and without such collections it is hard to perform comparable tests of alternative interfaces and improvements. Without a thorough, baseline description of the artefact created during the process of managing files, it is difficult to model that process and the many possible determinants suggested by past studies (e.g., technological, demographic, and individual differences).

Here we examine the contents of 348 collections used for personal, work, and study activities. We provide tabular data that describe the typical composition of such collections, discuss notable differences across collection types, and examine potential causes of content duplication. We end by discussing implications for future work.

**LITERATURE REVIEW AND RESEARCH QUESTIONS**
Personal information management refers to the study and practice of individuals managing information owned by or about them *and* to individuals personally managing information, owned by them or otherwise (Jones, Dinneen, Capra, Pérez-Quiñones, & Diekema, 2017), typically with the intention of later returning to that information. PIM research has explored a broad range of contexts and activities within this scope, including the common computing phenomenon of managing digital files, where personal could mean, for example, using a private computer to backup travel photos, using a company computer to organise project files, or retrieving manuscripts previously downloaded to a laboratory computer. Beyond PIM, files have been considered important digital possessions (Cushing, 2013) and a common part of personal archives (Marshall, 2008).

Hundreds of studies have covered various topics in file management (Dinneen & Julien, 2019), like searching and navigating to files (Bergman, Tene-Rubinstein & Shalom, 2013; Teevan, Alvarado, Ackerman, & Karger, 2004), file sharing (Capra, Vardell, & Brennan, 2014), and developing augmentations to FM software (Fitchett, Cockburn, & Gutwin, 2013). To investigate the common PIM activities of storing, organising, and retrieving (or exploiting) digital items (Jones *et al.*, 2017; Whittaker, 2011), many studies have characterised the artefacts produced by FM activities: people's digital collections (i.e., files and folders). Specifically, studies have measured collections' *scale* (i.e., how many files users store; Boardman & Sasse, 2004; Gonçalves & Jorge, 2003), *structure* (i.e., how files and folders are organised; Bergman, Whittaker, Sanderson, Nachmias, & Ramamoorthy, 2010; Henderson & Srinivasan, 2009), and *contents* (i.e., files and their types, discussed next).

**Contents of file collections**
The contents of file collections (i.e., the files a user manages on some computer) are generated or acquired by users actively storing files and folders (e.g., by creating and downloading them) or passively keeping (i.e., not deleting) files and folders (e.g., in downloaded archives; Watkins, Sellen & Lindley, 2015). Users' files come from various sources, including the Web (Jones, Bruce, & Dumais, 2001), external devices (Capra, Vardell, & Brennan, 2014), and peer-to-peer or cloud software (Marshall & Tang, 2012). Users keep files for many different purposes (e.g., personal or professional) and reasons, for example, because they are needed only briefly, or are being modified over the duration of some ongoing project, or will be needed again at some future time (Nardi, Anderson, & Erickson, 1995). In extreme cases, participants self-identifying as hoarders and minimalists have explained that they hoard because they are afraid of missing some data at a crucial moment, or do otherwise because they are reacting to or trying to prevent information overload (Vitale, Janzen & McGrenere, 2018). The end result is that, even in typical cases, collections contain tens to hundreds of thousands of files across tens of thousands of folders (Dinneen *et al.*, 2019), and those items necessarily compete for attention when users navigate through and retrieve them. It is therefore hard not to wonder: what are the contents of these collections?

Beyond satisfying scholarly curiosity, answering the above question also has practical implications. Evaluating PIM prototypes (e.g., those designed to improve user experience) requires test collections, and making sound comparisons across prototypes requires consistency in those collections (Gurrin *et al.*, 2019). However, no such collections are available for testing FM prototypes, as was acknowledged a decade ago (Chernov, Demartini, Herder, Kopycki, & Nejdl, 2008). A thorough description of collections' contents is also needed for studying and modelling the many possible determinants suggested by past studies, including technological (Bergman *et al.*, 2010; Boardman & Sasse, 2004), demographic (Khoo *et al.*, 2007), and individual (Lansdale, 1988) differences, and to compare the FM habits and collections of particular professional or social groups to a baseline, but such modelling, study, and comparisons require a baseline of data to compare to.

Five recent (i.e., in the last two decades) studies have sought to understand the contents of people's collections, evidenced by file extensions or categorised file types. Two studies of work collections – one of eleven varied professionals (Goncalves & Jorge, 2003) and one of seventy-three university staff (Henderson, 2009) – mostly agree in their findings: collections are, on average, approximately 30% text files, 20% image files, 10% unidentifiable, with spreadsheets, presentations, system files, source code, and extensionless files each comprising less than 10%, and audio, video, shortcuts, and other kinds of files being relatively negligible (i.e., approaching 0%). A third study, of forty engineers' collections (Hicks *et al.*, 2008), differed in that it showed 10% less text and a considerably higher 24% unidentified files. Finally, two studies (Agrawal, Bolosky, Douceur, & Lorch, 2007; Douceur & Bolosky, 1999) observed thousands of Microsoft employees' work collections over six years, but reported the thirty most common file extensions rather than categorised file types, preventing easy comparison to the other previously mentioned works. They observed .gif files to be the most frequent (9%), *.h*, *.htm*, and *.dll* files to be slightly less frequent, and *.jpg*, *.bmp*, and *.doc* files to be infrequent (e.g., 0.6% for *.doc*), at least until extensionless files became more frequent than any single extension. However, the authors note that the presence of technical staff collections, which make up the majority, have skewed the results as some differences were present among job roles (e.g., technical staff had considerably fewer *.gif* and more *.c* files). Such differences are intuitive – it seems reasonable to assume developers keep programming files, while teachers keep presentations and documents – but while such differences have been suggested by past works (Khoo *et al.*, 2007; Kwaśnik, 1991), they have not been explored or quantified.

**Duplication in file collections**
An additional concern about the contents of file collections is content duplication (i.e., file and folder duplication). Duplication can cause confusion during navigation and search, sharing, and version control, and duplicates may occupy limited storage space. Indeed, in one study 47% of participants reported struggling with duplicate files (Henderson, 2011). Several tools have been developed to prevent and treat duplication, for example file version control and software to identify duplicate images (Bergman, Tucker & Dahamshy, 2018), but is unclear if such tools are commonly used, and duplication nonetheless persists; additional potential causes therefore warrant consideration.

Duplication has, so far, been studied by examining the presence of repeating file and folder names, both of which appear common, ranging from 20% (Henderson, 2005, 2011) to 30% (Hicks *et al.*, 2008) duplication in collections. The causes of duplication are unclear; participants often suggest for example, they may download the same file from an email multiple times (Henderson, 2011). But given the size of collections and the high percentage of duplication, it is unclear if incremental actions like downloading attachments could be the primary cause. Surprisingly, only four participants suggested copy and pasting folders may lead to duplication, but recent works suggests this may be a primary cause, as multiplicative actions (like copy and paste) may shape other attributes of the file system (like extreme imbalance in file classification; Dinneen *et al.*, 2019). Some telling properties of duplication have been examined; for example, file name duplication has been seen to correlate strongly with folder name duplication, collection size (in files and folders), and the mean depth of folders within the folder tree (Henderson, 2011), but additional file system variables possibly related to duplication remain to be examined (e.g., file classification skew and the maximum depth of the folder tree) or the time required to produce duplicates (i.e., the age of the collection).

**Summary and research questions**
In short, we do not know what is in personal file collections, how work collections differ, or what causes content duplication, and these gaps in knowledge limit our ability to model FM, test improvements, and support computer users. To address these issues, in this manuscript we examine 349 collections to answer the following research questions:

RQ1. What are the typical contents of people's personal, professional, and scholarly file collections?

RQ2. How do the contents of different collection types differ?

RQ3. Do file and folder duplication correlate with other collection traits (e.g., folder depth or classification skew)?

**METHODOLOGY**
**Recruitment and data collection**
Various methods have been used to collect data about personal file collections, such as guided tours of users' computers (Thomson, 2015), recording participants desktops during structured tasks (Bergman *et al.*, 2010), and taking snapshots of user's collections (e.g., Henderson, 2009). We used the last approach, employing cross-platform, open-source software called Cardinal, described and validated by Dinneen, Odoni, Frissen, and Julien (2016) and used later by Dinneen *et al.* (2019).

When run, Cardinal solicits a participant's demographic information and use(s) of the computer, then accesses the portions of their folder hierarchy where they have specified they manage files, and records various properties (e.g., file extensions). Known system folders were ignored unless the user explicitly stated they manage files within them, and hidden files and folders were assumed to be either actively hidden by participants (i.e., we are not meant to see them) or else unknown to participants (i.e., not managed by them), and so were ignored. Files pointing to other files (i.e., shortcuts, aliases, and symlinks) were counted as pointers rather than the kind of file they pointed to. Though no file or folder names were recorded, instances of non-unique names were noted to provide a measure of duplication.

We recruited participants[1] from February of 2016 to August of 2018 by posting calls on study recruitment Websites, online communities (e.g., Facebook, Reddit), and mailing lists (e.g., industrial, governmental, and academic). Criteria for participation were only that participants have files stored locally (i.e., not exclusively in the cloud, whether at home, work, or school) that they personally manage, and have the abilities to read English and download and run the software. Participation thus consisted of running the software, answering the questionnaires, specifying where on the computer they manage files (both active, working areas and backup locations like external drives were encouraged), and reviewing a summary of the results before choosing to let the software submit the data to the researchers. Participation was therefore remote and anonymous.

**Data analysis**
To facilitate comparison with prior studies, file extensions were categorised into file types. There is currently no common taxonomy of file types nor consensus about how to categorise extensions, but the skewed distributions of extension frequency previously observed (Douceur & Bolosky, 1999) suggest that categorising the few, most common extensions will account for most files. We adapted the 18 file type categories used by Goncalves and Jorge (2003), with some modifications: (1) the *source code* and *Web script* categories were merged into a new category, *development* files; (2) *shockwave* was merged into *video*; (3) *PDA-related* was merged into *PIM*; and (4) a *pointers* category was added to account for shortcuts and symlinks. The result was 15 categories, discussed below, which have only minor differences to the categories used by Goncalves and Jorge (2003), Henderson (2011), and Hicks *et al.* (2008).

---

[1] This study was approved by ethics committees at Victoria University of Wellington (HEC #25658) and McGill University (REB #75-0715).

We categorised the most common extensions within each participant's collection until each collection was at least 80% categorised (excluding two outliers) and unidentified files comprised on average only 7% of a collection. This entailed categorising 453 extensions, each of which we researched to determine the best category, and which together include all the extensions listed in the prior works discussed above. The extensions, categories, and python function to categorise them are available on GitHub.[2]

Collections were divided into those used for personal matters, work, or study, according to participants' responses. To account for potential differences in work collections (discussed above), those were further divided (using participants' stated occupations) into three broad groups: (1) knowledge workers, (2) IT staff, or (3) other. Knowledge workers included quality assurance managers, doctors, journalists, librarians, and so on, whereas IT staff included positions like programmer, system administrator, or IT technician. The *other* category included occupations that are not obviously knowledge work (e.g., tradespeople, retail associates, and artists), null values, and people unemployed at the time of participation.

Following Henderson (2011), proportions of file and folder duplication within each collection was inferred by the presence of duplicate file and folder names. Limitations to this approach are discussed below.

As in prior studies of computer files, data along most measures were highly skewed, and approximated log normal distributions even after outliers were removed using interquartile range (Wilcox, 2011). Thus, to convey the *range* of typical values and avoid overestimation, log normal medians and means – hereafter, median* and mean* – are reported when appropriate (see Limpert, Stahel, & Abbt, 2001 for more detail). Further, nonparametric tests were used: Mann-Whitney U tests assess significance ($p$) in differences between groups with skewed data of unequal variance, and Kendall's tau ($\tau$) assesses correlation strength. Though such tests are often used for ordinal (rank) data, they perform well with continuous (interval or ratio) data and are more suitable for heavily skewed distributions than parametric equivalents (e.g., t tests and Pearson's rho; Chok, 2010). Differences between groups were tested pairwise (with the Mann-Whitney U test) since *n*-way tests like MANOVA require normal data distributions and log transformation of skewed data containing many 0 values is not straightforward.

**RESULTS**

We received data describing 348 collections, totalling 50 million files and nearly 8 million folders. Table 1 summarises the samples' demographic features, uses of the collections, and operating systems.

| Sample trait | Values |
| --- | --- |
| Age | Range 14-64<br>Mean 30 (SD 9.96) |
| Gender | Male: 218 (63%)<br>Female: 123 (35%)<br>Other: 7 (2%) |
| Operating system | Mac OS: 169 (48%)<br>Windows: 135 (39%)<br>GNU/Linux: 44 (13%) |
| Collection use | Work: 166 (48%)<br>    Knowledge work: 93<br>    Other: 48<br>    IT: 25<br>School: 143 (41%)<br>Personal: 39 (11%) |

**Table 1. Summary of attributes of sampled participants, systems, and collections.**

Complete statistical descriptions of each collection type are provided on the Web[2], while in this section we present an overview of the range of file extensions, the categorised collections, how collection types contrast, and results of inferential testing of duplication and related phenomena. For brevity, after describing personal collections we present only the strongest differences between collection types rather than describing each exhaustively.

**File extensions**

The collected data contain 85,000 unique extensions, distributed in a highly skewed and long-tailed manner as observed in prior works (Douceur & Bolosky, 1999). Thus, despite our categorisation efforts accounting for less than 1% of the variety of extensions, and despite extensionless files comprising on average 10%, 83% of all files were categorised and on average only 7% of each collection was left uncategorised.

**Personal collections**

Personal collections were, in typical cases, comprised of 45 thousand to 85 thousand files across 7 thousand to 21 thousand folders. The data across most file types was highly skewed, indicating great variation across the collections, but a few general claims can be made. Images and development files were the most common types at 22-29% and 4-35%, respectively (e.g., 20,000 image files and 12,000 development files, depending on collection size). Next most common were system, audio, and extensionless files at roughly 5-25% each (e.g., ~7,000 of each kind). The typical proportion of audio files varied greatly, however, from 1% to 20%. Unidentified files comprised 6-9% of collections, and text files 3-5% (e.g., ~2,000 text files). PIM and video files were relatively scarce (0-1%), and every other type (i.e., databases, spreadsheets, executables, backups, presentations, bookmarks, and pointers) was relatively non-existent (e.g., a dozen or so files each).

---
[2] https://github.com/jddinneen/file-extension-categoriser

**Differences across groups**

Figure 1 displays treemaps of example personal, knowledge work, and IT collections that feature proportions of file types within the typical range (e.g., median* and mean*) for their respective groups. Treemapping is a method for visually displaying differences in group compositions, and originally, file system data (Shneiderman & Wattenberg, 2001). The full results tables should be consulted for discrete values and comprehensive descriptions of the composition of typical collections in each group.

On average, personal collections were smaller than all other collection types while IT collections were the largest, generally ten times larger than personal collections ($p$=0.059) at 61 thousand to 829 thousand files. Knowledge workers' collections had the least folders, typically 2.5 to 14 thousand, or about half that of typical student collections ($p$=0.006) and a fifth of IT collections ($p$=0.007).

Most collection types had proportions of images and development files comparable to personal collections, with the most notable differences being that on average IT collections had fewer images (11-26%; $p$=0.019) and more development files (up to 49% even in typical cases; $p$<0.001). Knowledge workers also had smaller proportions of images than personal collections (16-24%; $p$=0.076).

Personal and *other* work collections had the most system files (5-24%), whereas other groups had fewer with IT staff having the fewest (2-5%; $p$=0.0495 when comparing to other). Personal collections contained the highest proportion of audio files of all collection types, with students having the second most and knowledge workers' collections having the least (0-1%; $p$<0.001 when comparing to students). Although IT collections resembled personal collections in the proportion of text files, knowledge workers', students', and other workers' collections each contained almost twice as many (5-10%; $p$<0.001 when comparing knowledge workers' to personal collections). In all collection types there were relatively few video files (e.g., 0-2%) and almost none (approaching 0%) of all remaining file types; for example, even knowledge workers had ~0% presentation files.

All work and student collections exhibited smaller proportions of extensionless files than personal collections did, and with much narrower ranges of typical values, but the pairwise differences were not statistically significant; for example IT collections had on average 4-6% (SD* 2.3) as compared with 2-29% in personal collections (SD* of 9.7; $p$=0.359).

**Duplication results**

Across all groups we observed that 23-34% of files in a collection were typically duplicated (excludes two outliers), while on average 44% (traditional mean) of folders were duplicated. IT collections exhibited the most duplication of files (40-48%), and personal collections the least (20-26%; $p$=0.002 when comparing to IT), while folder duplication was greater overall but followed a similar pattern: IT collections had the most (53-60%), and personal collections had the least (33-47%; p=0.021).

All correlations reported here were statistically significant ($p$<0.001; omitted below for brevity). File and folder duplication correlated moderately (τ=0.39), and each correlated moderately with their total counts (τ=0.36 for files, 0.40 for folders). We found weak correlations between folder tree width and file and folder duplication (τ=0.29 and 0.33, respectively), and folder depth correlated moderately with file duplication (τ=0.45) and folder duplication (τ=0.48), and similarly, maximum folder tree depth (i.e., the deepest part of the tree) correlated moderately with folder duplication (τ=0.45) and weakly with file duplication (τ=0.26). We also found weak correlations between skewness in the classification of files into folders (i.e., the degree to which files are placed into relatively few folders, leaving most folders relatively empty) and file (τ=0.16) and folder duplication (τ=0.29). Duplication had no linear relationship with the age of collections whether measured by mean or maximum file age (e.g., τ < 0.05).

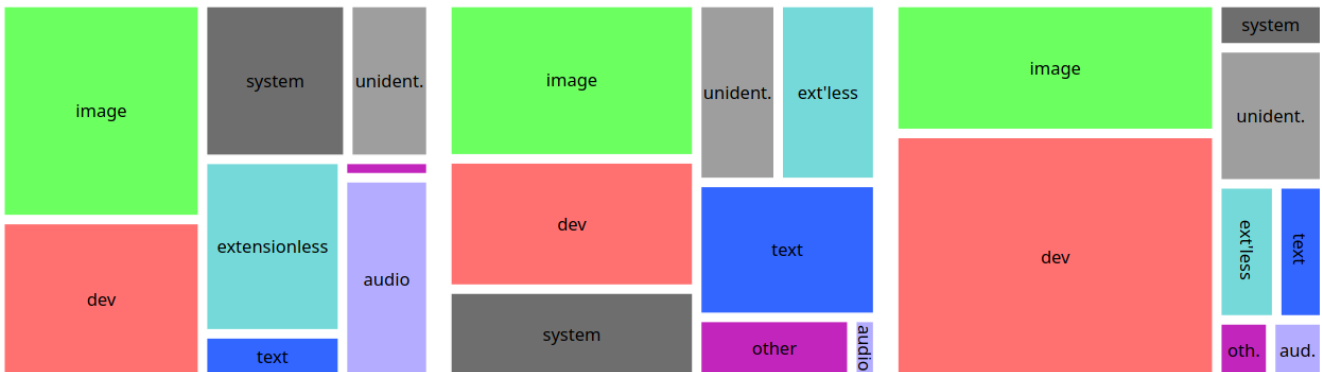

**Figure 1. Treemaps showing examples of typical collections for personal use (left), knowledge work (middle), and IT work (right). All collection types typically contain many images and development files and relatively few of the rarest file types (e.g., video, databases, etc; here collectively called *other*). Personal collections contain more images and audio files, knowledge workers' collections contain more text files, and IT collections contain more development files and few system files.**

## DISCUSSION

Here we synthesise and discuss our results and compare them to commensurable findings from prior works.

### File extensions

We saw 85 thousand extensions, a far greater variety than reported in prior works (e.g., 623 occurring more than once; Henderson, 2009). This suggests file extension variety is growing, but the distribution of extensions remains a similar shape (e.g., following Lotka's law; Douceur & Bolosky, 1999) and consequently categorising the majority of the collection can still be done with relatively few (e.g., 450) extensions (c.f. 350; Goncalves & Jorge, 2003).

### Characterisations of collections by type

RQ1 asked "what are the typical contents of people's personal, professional, and scholarly file collections?" The tabular data provide an uninterpreted answer to the question, which we contextualise below by characterising and interpreting each collection type, with emphasis on differences between collection types so to answer RQ2.

*Personal collections* are the smallest (in terms of files), and despite containing relatively many image files (as did most collection types) they exhibit some uniqueness in their composition. They had the most audio files and system files, second most development files (behind IT), and widest range of extensionless files. While image and audio files are unsurprising given that many people keep personal music and image collections, the prevalence of system files is surprising, particularly as the data collection software scanned only user-managed folders and ignored system folders unless users specified otherwise. It is similarly unclear what caused the presence of so many development files and the occasional cases of so many extensionless files.

Personal collections also exhibited the widest range of file types, perhaps reflecting a wide variety personal computing activities that may not be reflected in PIM research focusing on professional contexts. Surprisingly, like most other collections, personal ones contained relatively few (e.g., 0.1%) file types like backups, executables, PIM files, and so on, thus suggesting the variety of file extensions are variations within the main categories (e.g., different image extensions). Personal collections also featured relatively few text files, likely reflecting the popularity of such formats (e.g., *.docx*) in knowledge work (as contrasted with personal matters). Perhaps because of their size (considered below), personal collections featured the least duplication, though still roughly 20-50%.

*Knowledge workers' collections* (e.g., work collections managed by doctors, teachers, librarians…) were the smallest of the work collections, and while bigger than personal collections, they nonetheless featured ~30% fewer folders, which is surprising given the popularity and many uses of folders identified in previous PIM research on knowledge workers. Such collections had the least audio files, perhaps indicating an absence of music collections on work machines, and predictably, more text files, perhaps due to reasons considered above.

The percentages of system, development, and images files observed in knowledge workers' collections are roughly consistent (i.e., within 2-3%) with prior works (Goncalves & Jorge, 2003; Henderson, 2009; Hicks *et al.*, 2008), suggesting no meaningful change over time, but the percentage of text files observed (7-12%) was much less than in those works, which saw 21-34%. The cause for this difference is unclear; given the data are highly skewed, it is possible that in prior works averages were inflated by outliers. That knowledge workers had relatively few archive, audio, video, shortcut, PIM, and database files also roughly aligns with the prior similar studies, but surprisingly, knowledge workers kept relatively few (e.g., < 1%) of other types that are commonly associated with knowledge work, like spreadsheets and presentation files, which contrasts with those works, which found 4-8% spreadsheets and 2-7% executable files. Such differences may be attributable to the high portion of previously uncategorised files (e.g., 24%) or to occupational differences (e.g., examining only engineers; Hicks *et al.*, 2008).

*Other work collections* were generally big, but otherwise like the other work collections in most ways. However, they did feature more image and audio file percentages, comparable to personal collections, perhaps reflecting the multiple uses of machines people working in less traditional roles (e.g., whereas knowledge workers may not store many personal images on machines owned by their employers, freelance designers may work from the same machine on which they store such photos). These collections were also highly varied, which may reflect the diversity of the participants' occupations.

*Students' collections* resembled knowledge workers' collections, but with twice as many folders and more audio files; this may reflect that many students are knowledge workers in training (and/or that education is often knowledge work) and therefore may manage computer files similarly.

*IT collections* are huge and unsurprisingly contain many development files. They also contain relatively few other text files, but surprisingly, contain far fewer system files than other collections (and perhaps far fewer extensionless files, though this lacked statistical significance). Absolute values indicate the difference is not simply due to having more development files (i.e., thus lowering the proportion of system files): IT collections contain up to 5,500 system files, whereas knowledge workers' collections typically contain up to 32,000. Although in this manuscript we do not examine differences across operating systems, because IT collections were distributed no differently across the OSes than other collections (and comprised only 16% of the Linux machines) it seems unlikely OS is the main cause of the relative scarcity of system files in IT collections. It is possible that IT professionals make a concerted effort to separate their systems' files from other work files (and/or consider it routine maintenance), whereas, for example, it may be rare for knowledge workers to do so. That IT collections had 0% executables shows a decrease from 3% in 1999 (Douceur & Bolosky), and likely reflects general collection growth (e.g., in development files) over time.

RQ2 asked "How do the contents of collection types differ?" In summary, personal collections are generally smaller than work collections and contain more images and audio files, and are generally more varied than other collection types. Knowledge workers' collections contain more text files and far fewer folders than personal collections. IT collections exhibit some predictable traits like having many development files, but also some surprising ones, like having far fewer system files.

**File and folder duplication**
Results regarding duplication were broadly similar to but weaker than observations made by Henderson (2011), with the most notable exception being that we observed more duplication overall: roughly 30% for files (matching Hicks, 2008) and 44% for folders (as compared with ~20%). This may be attributable to general collection growth over the last decade – Henderson (2011) suggests that larger collections may have more duplication, and the results here support this – but the results here also suggest the story is more complicated. Specifically, the presence of several correlations (ranging from weak to moderate) between duplication and structural variables suggests that multiplicative actions may not just cause duplication directly (pasting causes duplicates), but also *indirectly*: multiplying parts of the tree produces a more complex structure that is harder to navigate (Julien *et al.*, 2013), which is therefore more costly to maintain. For example, users may multiply branches of the tree *downward*, thus increasing measures of file system depth, and consequently be less likely to later navigate back down to those files to address the resulting duplicates. This could explain why measures of depth are the strongest correlates with duplication, stronger even than collection size.

Surprisingly, we found no notable correlations between duplicate files or folders and the age of collections whether measuring age by the mean age of files or the oldest file in a collection. This is surprising because prior studies (e.g., Dinneen *et al.*, 2019; Henderson, 2011) suggest that it takes time for a collection to grow, even multiplicatively, and thus older collections should exhibit greater duplication. It is possible the majority of duplication is produced near to the creation of a folder system and that only incremental duplication (i.e., downloading duplicates) happens later, though this warrants further investigation.

RQ3 asked if file and folder duplication correlate with other collection traits. In summary, duplication does appear to correlate with structural properties of the file system, especially folder depth. Given that duplication is so common despite the availability of anti-duplication tools like those mentioned above, perhaps additional strategies should be implemented to prevent and treat duplication; for example, upon a user initiating a large copy and paste action, the file manager interface could provide a dialogue to exclude some subfolders, and the OS could periodically scan for duplicates and prompt the user to resolve them.

**Implications for assessing collections**
The tabular data and characterisations of collections provided by this study can be consulted when assessing existing collections managed by particular groups of participants, and also when generating or selecting collections for use in evaluations of PIM software. For example, a study examining digital information management of administrative staff at a university could compare the contents of their collections to the knowledge workers' collections described here to assess if the participants are representative or typical (e.g., if those collections have <7% or >12% text files, they are not typical). Similarly, to increase the generalisability of evaluations of prototypes like file manager plugins or desktop search improvements, the collections used should be representative of those likely to be used in practice (e.g., in personal contexts, or by IT staff, or by general users), and this can be achieved by consulting the ranges of typical values reported here and choosing or generating only collections that fit within the ranges given.

## LIMITATIONS

The present study has notable limitations, some of which are attributable to using extensions to identify file types: 23% of all files did not fit a traditional file type category, either because they had no extensions or had extensions so rare we did not categorise them (i.e., 84.5k of the 85k different extensions, or 99% of the variety). Categorisation is often imperfect, and some extensions fit multiple categories (e.g., *.py* is a text file, used for development, and used by the system). While the majority of extensions fit neatly into one category, studies categorising files (and extensionless files) differently may observe different results. As our categorisation method is the most comprehensive thus far, and is shared on the Web, we hope it facilitates consistency and comparison in future works.

Similar limitations are present in the collection groupings. For example, though work collections were separated from those used only for personal matters, they included collections used for both, and those used strictly for work may differ further still. Likewise, the occupation categories used (e.g., knowledge workers) here are broad; future work may find it useful to examine differences across more specific title groupings, like those provided by the International Standard Classification of Occupations.

The method used in this study and prior works to infer duplicates from name duplication is imperfect as it can detect false positives (files can share names without containing the same content) and miss duplicates (files can have different names despite containing the same content). The effect of such limitations is currently unclear, but future studies may address them by implementing purpose-built methods for detecting duplication (e.g., full-text analysis or computer vision).

Finally, the data collection method used in this study necessarily omits some information useful in interpreting the results, like users' reports of *why* they keep duplicates or particular file types, and the analyses used were intended primarily to explore basic potential differences between collection types. In future work it would be useful to complement such quantitative descriptions with qualitative reports and further analyses for nonparametric data.

## CONCLUSION

Supporting people in the daily task of managing information requires understanding many dimensions of the relevant phenomena, including, for example, what is in their file collections. In this manuscript we established the contents of people's file collections, including personal, study, and work collections of various kinds, and discussed commonalities and notable differences among the broad collection types. In doing so we identified an extreme variety in file extensions, provided an updated categorisation scheme for the extensions, and discussed remaining limitations to be addressed. Trends and differences in these results across prior studies and across time were also discussed, including those related to file duplication. We also discussed the implications of the results for PIM research, and provide the statistical tables to facilitate such works. Expanding on those implications, we next consider concrete implications for both narrow and broader research.

Starting narrowly, several aspects of the data remain to be analysed, for example of duplication across file types, occupations, or operating systems (Bergman *et al.,* 2010), and of individual differences that may pertain to FM, like age, gender, education, and individual differences. Going forward, follow-up work would benefit from more actively involving participants in data collection to add valuable context (present in many qualitative studies) to the numeric results, thus potentially overcoming persistent methodological debates about qualitative and quantitative approaches and yielding a more holistic and comprehensive understanding of the relevant phenomena.

More broadly, considering the role of files in other areas of information science is also likely to be fruitful, and is a natural fit. For example, information behaviour research that examines how people "seek, manage… and use information" in wider information ecosystems (Fisher, Erdelez, & McKechnie, 2005, p. xix) and related activities like filing, archiving, and organising collected information (Meho & Tibbo, 2003) matches very closely previous characterisations of what users do with files (Boardman & Sasse, 2004; Malone, 1983). We hope the contents of people's digital collections can be useful in considering how PIM integrates into broader information behaviour.

One area that may benefit from FM research is (digital) cultural heritage: it may interest cultural heritage institutions (especially digital ones) to know what is in people's collections, because the contents and (original) order of personal collections of relevant individuals influences what value can later be made of such collections (e.g., by libraries, archives and museums) and how those collections can best be preserved and accessed. Further, the way people collect and organise their personal and work collections (often with blurred boundaries between them) can have impact on areas where personal and collective meets, such as records management (because personal attitudes, practices, and habits are often mirrored in the workplace) and digital curation, for example, because the way files are curated from the point of creation directly affects their longevity and authenticity (Oliver & Harvey, 2016). When personal collections belong to people whose life and work has significance for cultural heritage sector, the way they organised, preserved and explained their collections will greatly influence the interpretation and long-term usability of such collections. Some works have considered files as part of the personal archive (Cushing, 2013; Kaye et al., 2006; Marshall, Bly, & Brun-Cottan, 2006), but fuller synthesis of the areas remains to be done.

In summary, while the present manuscript elucidates what is in people's file collections and what might cause duplication, we hope to have further stimulated researchers in and beyond PIM to consider the everyday role of FM.

**ACKNOWLEDGMENTS**

We thank our participants for their time and effort, and Maja Krtalić and anonymous peer reviewers for their feedback.